\documentclass[conference]{IEEEtran}
\usepackage{flushend}
\usepackage{pifont, comment}
\usepackage{times}
\usepackage{url}
\usepackage{latexsym}
\usepackage{stfloats}
\usepackage{mathrsfs}
\usepackage{lineno,hyperref}
\usepackage{longtable}

\usepackage{lipsum}
\usepackage{pdfpages}
\ifCLASSINFOpdf
\else
\fi
\title{A speech-based driver assisting module for Intelligent Transport System}

\author{\IEEEauthorblockN{Himangshu Sarma}
\IEEEauthorblockA{Department of Computer Science \& Engineering\\
Indian Institute of Information Technology Senapati \\Manipur, India\\
Email: himangshu.tezu@gmail.com}
\and
\IEEEauthorblockN{Navanath Saharia}
\IEEEauthorblockA{Department of Computer Science \& Engineering\\
Indian Institute of Information Technology Senapati \\Manipur, India\\
Email: nsaharia@iiitmanipur.ac.in}}

\begin{document}
\maketitle
\begin{abstract}Aim of this research is to transform images of roadside traffic panels to speech to assist the vehicle driver, which is a new approach in the state-of-the-art of the advanced driver assistance systems. The designed system comprises of three modules, where the first module is used to capture and detect the text area in traffic panels, second module is responsible for converting the image of the detected text area to editable text and the last module is responsible for transforming the text to speech. Additionally, during experiments, we developed a corpus of 250 images of traffic panels for two Indian languages.
\end{abstract}

\section{Introduction}
Detection and recognition of text in uncontrolled environment is complex due to blur panels because of speed, occlusion because of overtaking or other obstacles, deterioration of traffic panels because of age, and intensity of light. 
As a part of the intelligent transportation system, a number of work has been done over the years in detection and recognition of the traffic lights~\cite{gong2010recognition, john2014traffic, shi2016real}, recognition of the traffic signs~\cite{de2003traffic,fang2004automatic, paulo2007automatic,ellahyani2016traffic}, detection of the road boundaries~\cite{de2015automatic}, detection of the number of lanes~\cite{wujcicki2016automatic} and detection of the crakes in road~\cite{shi2016automatic,oliveira2014crackit}. 
Recognition of texts from traffic panels are also studied in the domain of image to text conversion~\cite{ellahyani2016traffic, gonzalez2014text,greenhalgh2015recognizing}.
This article describes an experiment to extract informations from images of roadside traffic panels and convert them to speech aiming to assist the vehicle drivers in the road by reading the traffic-panels loudly. 
The research concentrates on the following points. 
\begin{enumerate}

\item To detect and recognize the texts in traffic panels written in English and Indian languages. The extracted text will be passed through a speech generation module to assist the vehicle driver. 
\item To develop a Indian languages dataset of traffic panels for assisting the intelligent transportation researchers. Although, every state in India has more than one languages, literature survey hardly reveals reports on datasets of traffic panels, not even for the scheduled languages\footnote{https://en.wikipedia.org/wiki/Eighth\_Schedule\_to\_the\_Constitution\_of\_India; accessed date: 31 August 2018}.  
\item Being the second largest road-network\footnote{https://en.wikipedia.org/wiki/List\_of\_countries\_by\_road\_network\_size; accessed date: 31 August 2018} in the world such dataset may help in finding deteriorated or outdated traffic panels to the interested government agencies if the extracted images are monitored properly. Properly maintained repositories like traffic panels may help the driver or the automatic vehicle to take rapid decision, such as, which panel will appear next by mining the repositories or may work as an alternative of global positioning system. 
\end{enumerate}


Thus, the contribution of this research lies in development of a speech based intelligent transportation module and the traffic panel dataset developed for two Indian languages.   
In this research, texts of Assamese and Hindi languages were used as experimental test-bed with texts of English language, which was common language to both the traffic panels, where the script used by the languages are Assamese, Devanagari, and Roman respectively~\cite{saharia2014stemming, sarma2017development}. Thus, selection of the languages for this research was based on the script. The similarity among the considered languages tends to zero except loan words and named-entities.

Organisation of the paper is as follows. Reports related to the detection and extraction of text is discussed in Section II. Section III, discusses the working strategy of our image to speech conversion module. Experimental results, analysis and future direction are discussed in Section IV and Section V respectively.

\section{Literature survey} 
With the advent automatic vehicle and computer vision, advanced driving assistance modules such as traffic sign, light, and panel detection and recognition systems are attracting researches. Traffic-panel, which is an extended version of traffic-sign with texts~\cite{gonzalez2014text}, contains direction and distance to a location, and name of the path. Repository and processing modules of traffic-panels are important for urban areas, where the coverage of global positioning system is poor. 
We have added a speech generation module on the top of the text recognition module~\cite{ellahyani2016traffic, gonzalez2014text,greenhalgh2015recognizing} of traffic-panels. 
Greenhalgh et al.~\cite{greenhalgh2015recognizing} classified the problem of detection of texts from image into two main classes: region-based techniques that employ local features and connected component based techniques that employ global features such as colour density.
Recognition of words from images is complex because of the scenarios such as blurness and obstacles. Hidden Markov Model with Web Map Services was introduced to detect words from image by Gonzalez et al.~\cite{gonzalez2014text} to overcome the noises returned by the open-source image-to-text conversion module Tesseract~\cite{smith2007overview}. Chrominance analysis~\cite{}, Shi and Tomasi features~\cite{chen+05}, Fast-fourier transformation~\cite{reina+06} and Gaussian Mixture Models~\cite{chen+05} among other methods to detect lines and words from roadside panels. 

However, detection and recognition of traffic panels are still considered as a challenging problem due to issues associated with complexities in detecting and recognizing languages and texts, changing representation of symbols, texts, colours, and shape with changing region, and deviation of viewpoints. 

\section{Experiments}

\begin{figure*}[!htbp]

\begin{minipage}{.5\linewidth}
\centering
\label{main:Category 1}\includegraphics[scale=.1]{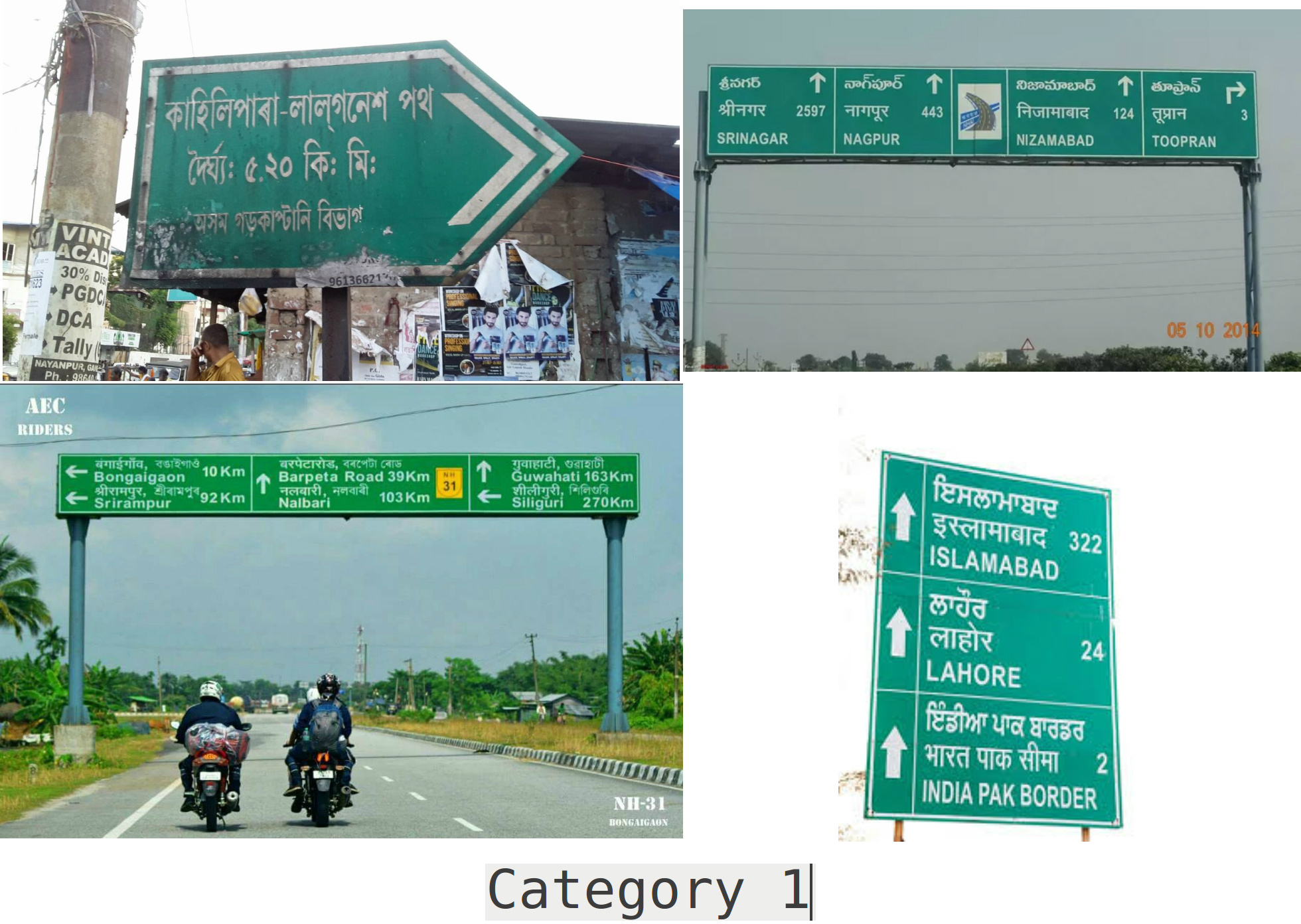}
\end{minipage}%
\begin{minipage}{.5\linewidth}
\centering
\label{main:Category 2}\includegraphics[scale=.1]{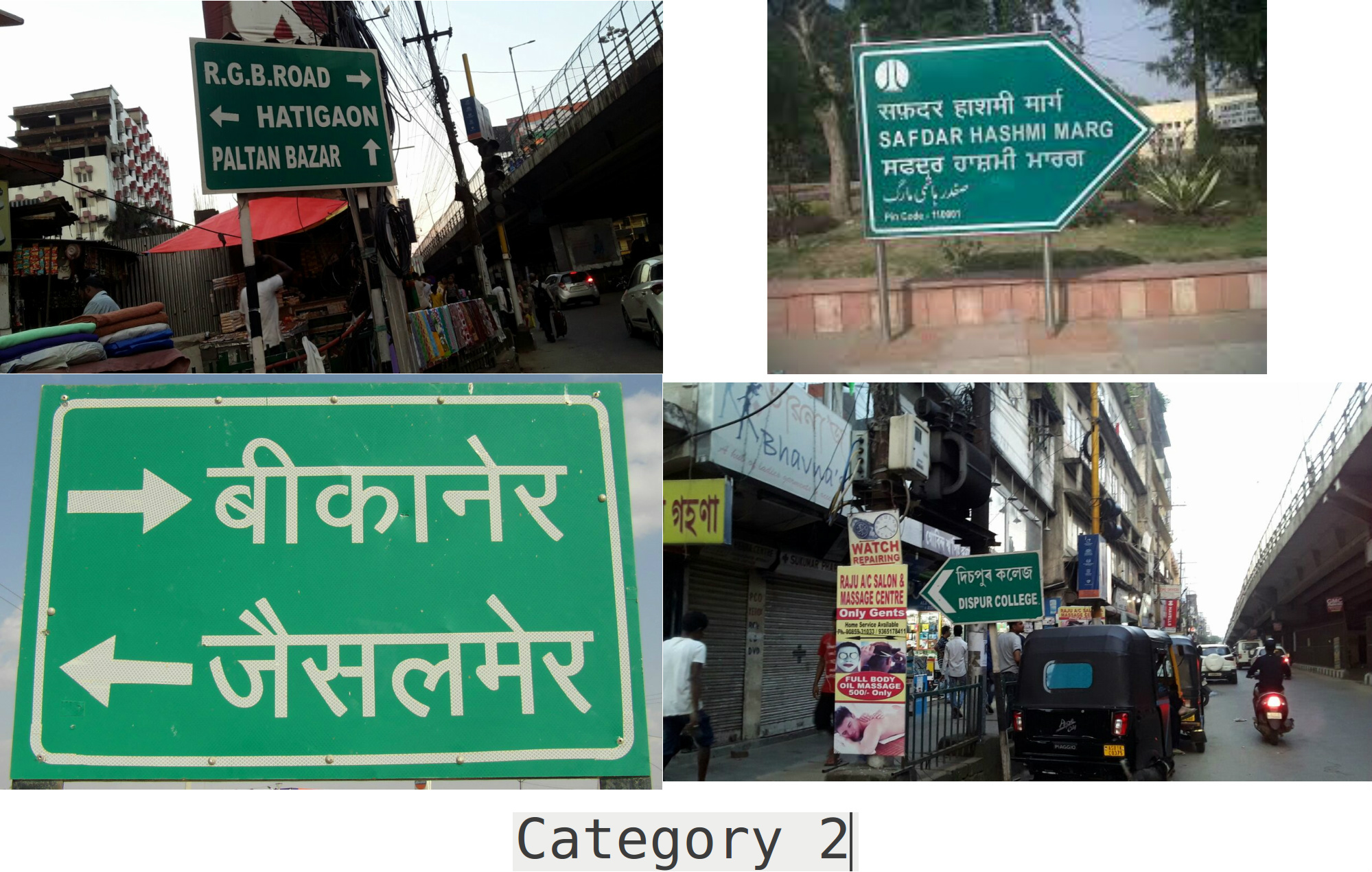}
\end{minipage}\par\medskip
\begin{minipage}{.5\linewidth}
\centering
\label{main:Category 3}\includegraphics[scale=.1]{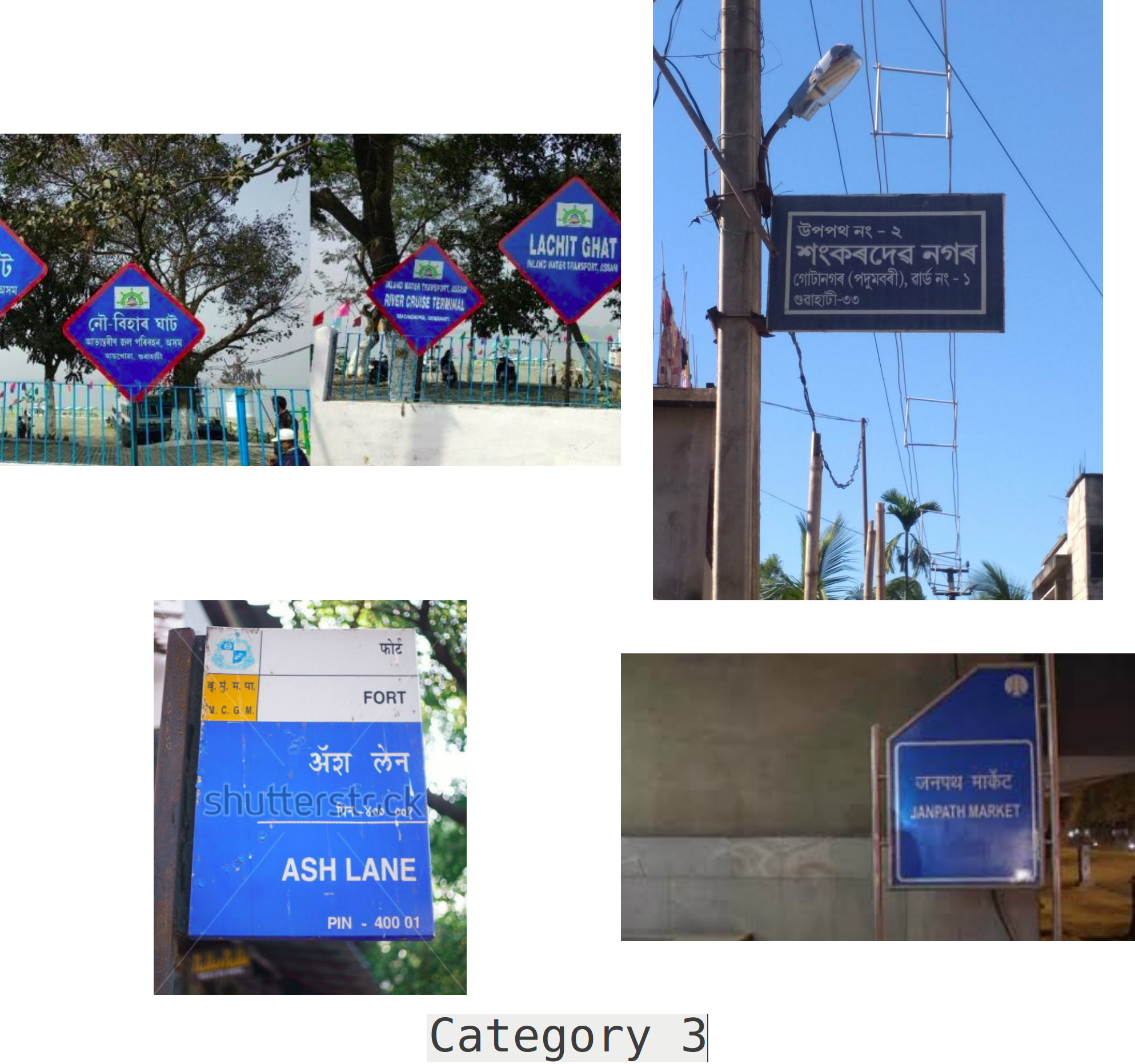}
\end{minipage}%
\begin{minipage}{.5\linewidth}
\centering
\label{main:Category 4}\includegraphics[scale=.1]{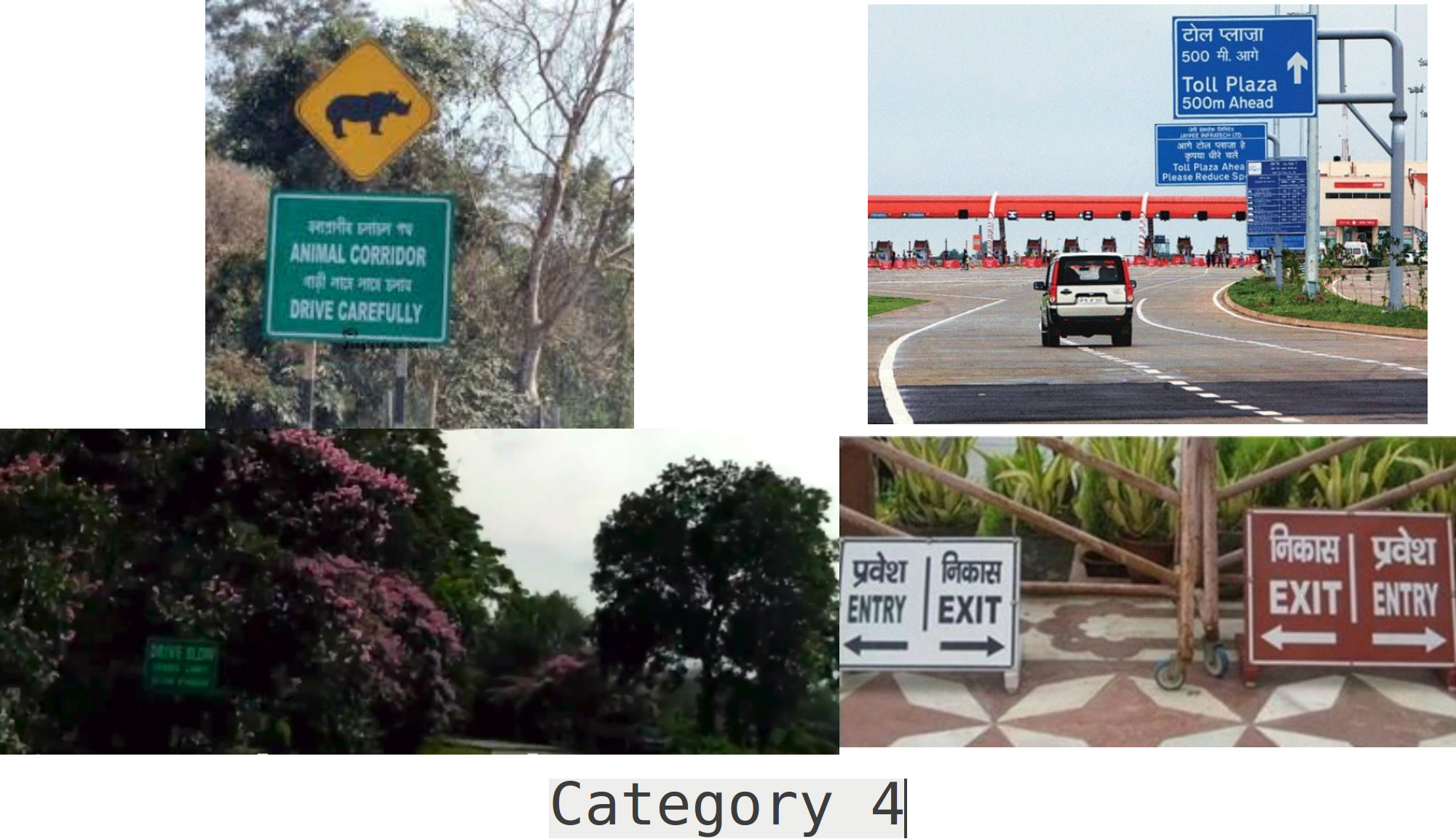}
\end{minipage}\par\medskip

\caption{Road panel Corpus}
\label{fig:main}
\end{figure*}

\subsection{Preparing the test-bed}
Though country-wise road-signs and road-panels are different, for this experiment we used road-symbols of Indian-subcontinent. In this article, we focus only on road sign or panels which contains different English and Indic languages text. We developed a corpus of around 250 road panels that contains Assamese, Hindi and English texts. Images are collected using the following approaches, they are:

\begin{itemize}
\item Extracting road panels from driving videos available online. 
\item Taking photographs personally from installed traffic panels. 
\item Collecting images of traffic-panels available online\footnote{which have been used only for research purposes}.
\end{itemize}

Also, we keep different categories of text based images in our corpus, they are:

\begin{itemize}
\item \textbf{\textit{Category 1:}} Images with destination name and distance
\item \textbf{\textit{Category 2:}} Images with the location name and the direction of the symbol
\item \textbf{\textit{Category 3:}} Images with the name of the location of that place
\item \textbf{\textit{Category 4:}} Cautious images with text, e.g., Drive slow, school ahead etc.
\end{itemize}

A set of images of the corpus is shown in Figure~\ref{fig:main} with different languages and categories.

\subsection{System architecture}

\begin{figure*}[!htbp]
\centering
\includegraphics[width=1.0\textwidth]{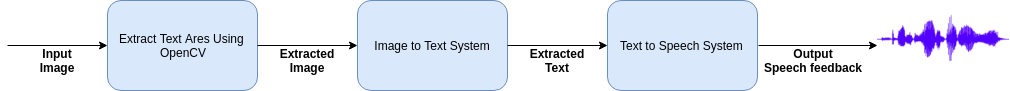}
\caption{System architecture}
\label{cys2}
\end{figure*}

Speech-based feedback to the drivers or the autonomous driver system is a very important application for the intelligent transport system.  The proposed system is mainly designed for speech-based feedback to the drivers for different kinds of road panels. The system will also very helpful for the autonomous driving.  An architecture of the system is shown in Figure~\ref{cys2}.

\begin{figure*}[!htbp]

\begin{minipage}{.5\linewidth}
\centering
\label{main:Category 1}\includegraphics[scale=.2]{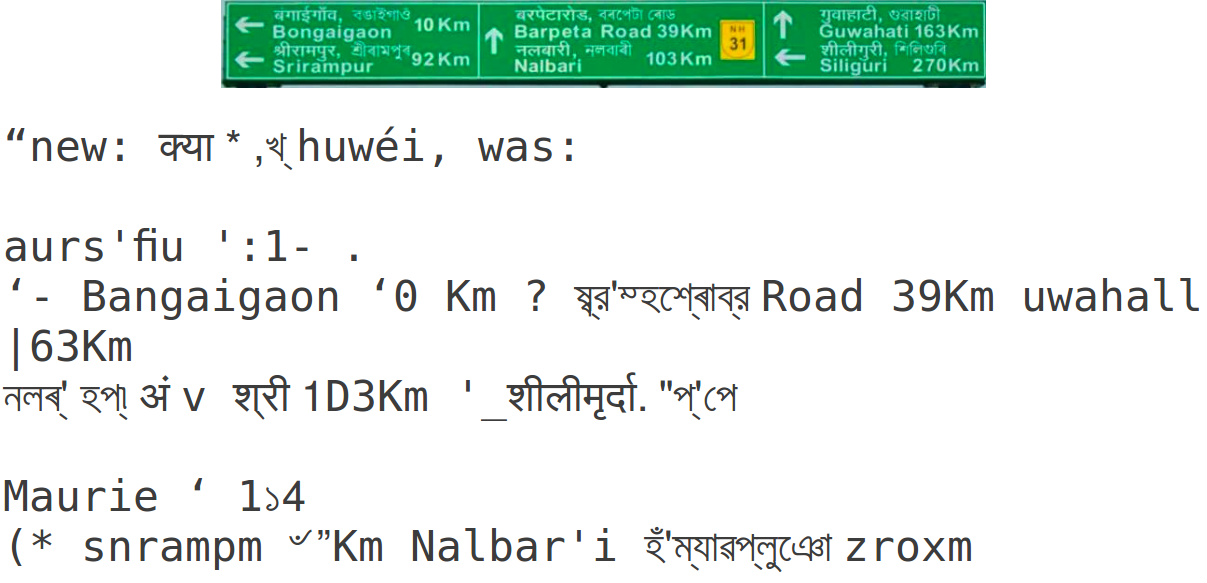}
\end{minipage}%
\begin{minipage}{.5\linewidth}
\centering
\label{main:Category 2}\includegraphics[scale=.2]{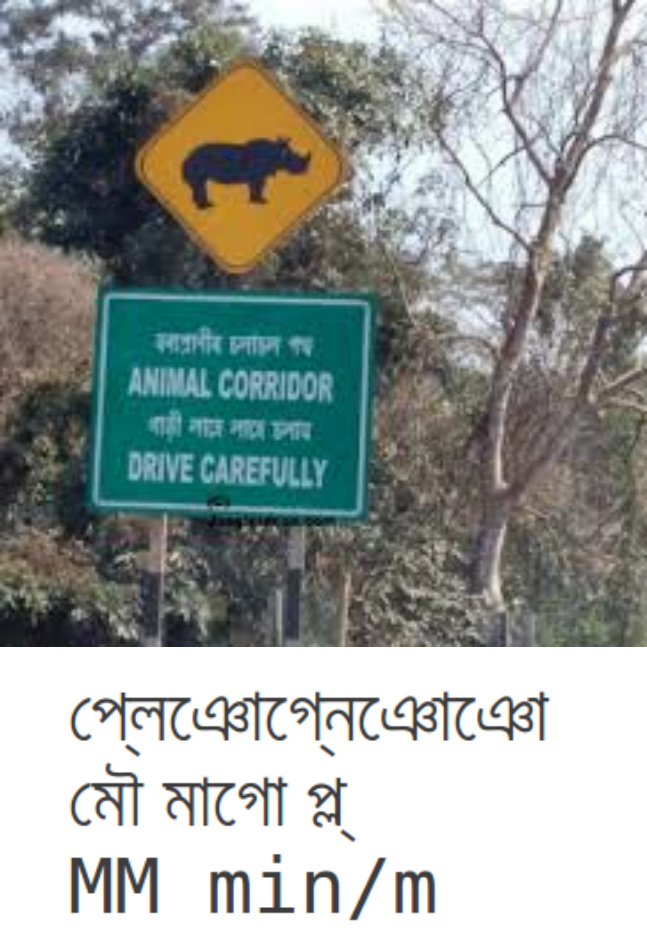}
\end{minipage}\par\medskip
\begin{minipage}{.5\linewidth}
\centering
\label{main:Category 3}\includegraphics[scale=.2]{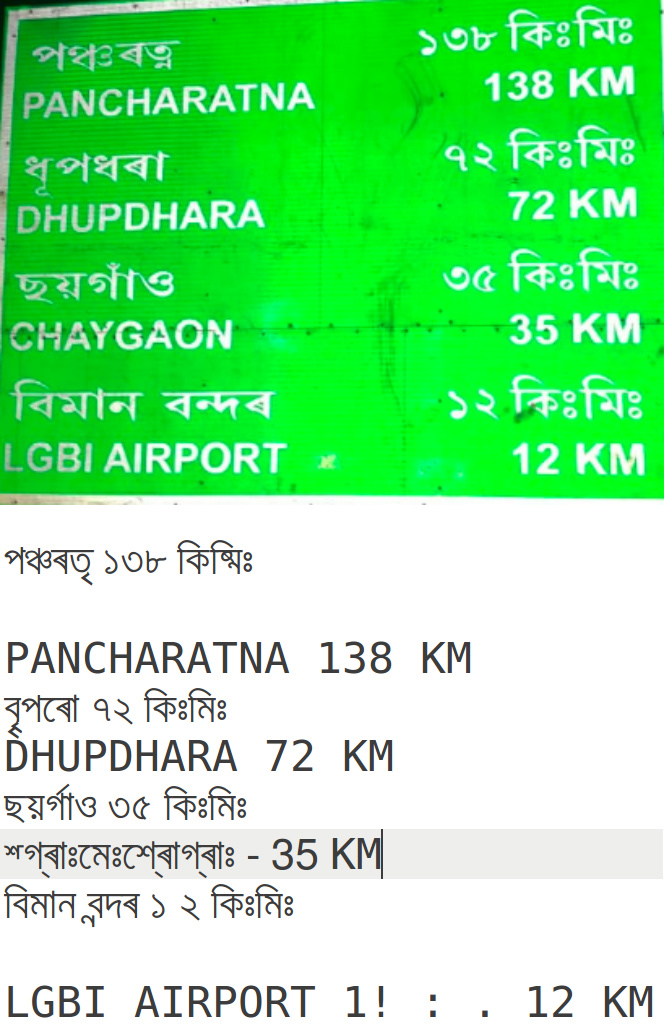}
\end{minipage}%
\begin{minipage}{.5\linewidth}
\centering
\label{main:Category 4}\includegraphics[scale=.1]{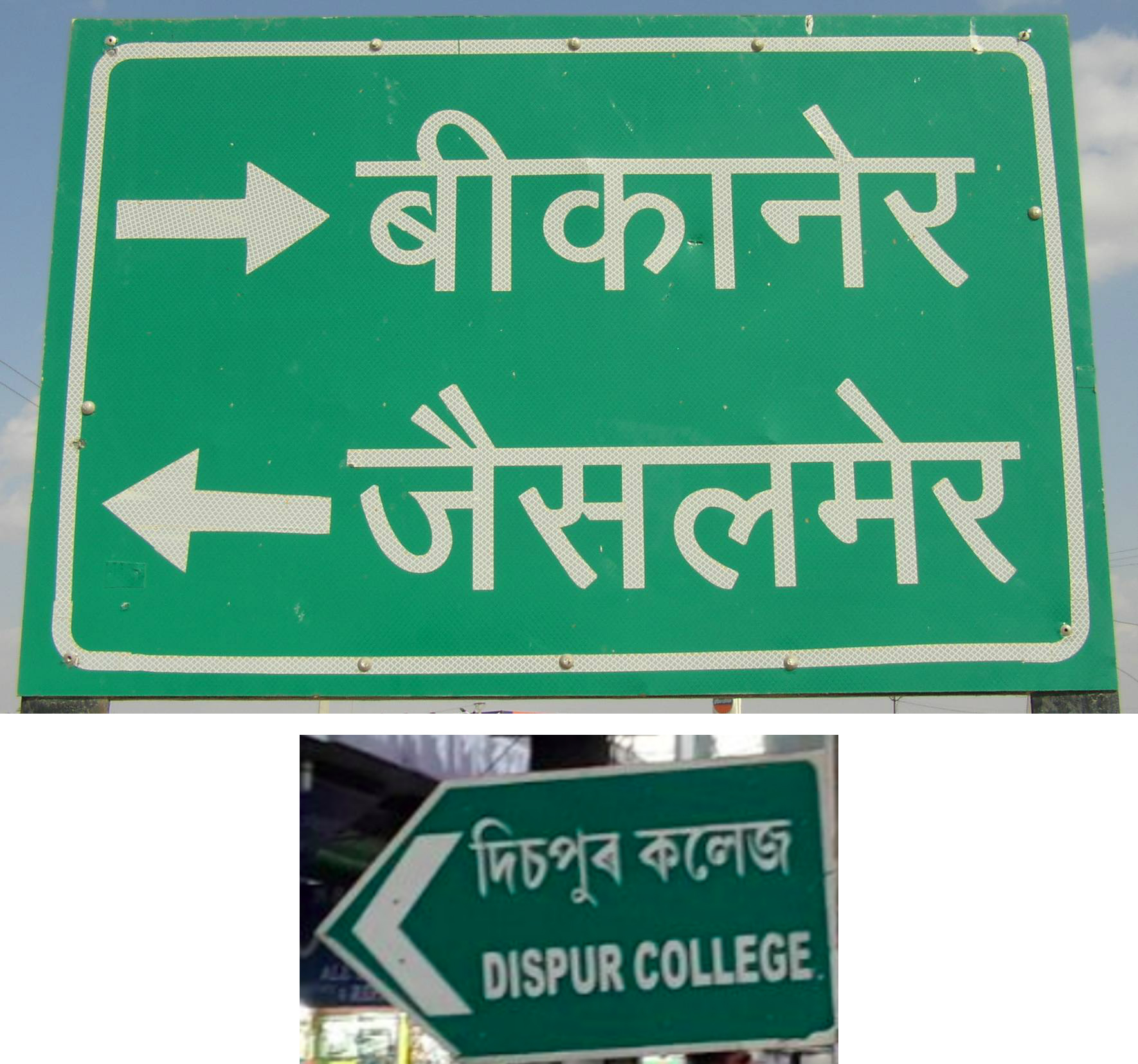}
\end{minipage}\par\medskip

\caption{Results of Image to text model}
\label{ocrRes}
\end{figure*}

The system consists of mainly three parts, they are:
\begin{itemize}
 \item \textbf{\textit{Extract text area:}} In this process, the system takes text-based road panel images as an input and extract the text areas of the images. To extract this in our system we used OpenCV. 
 \item \textbf{\textit{Image to text:}}  Here the system takes images of the text areas as an input. The system then goes through the image to text or OCR(Optical Character Recognition) system and extract the specific Indic languages, i.e., Assamese and Hindi along with English languages as an output in a text format. For this system we compare different OCR techniques, e.g., Tesseract~\cite{smith2007overview}, OCR.space\footnote{https://ocr.space/ ; Access Date: 30 August 2018} etc.
 \item \textbf{\textit{Text to speech system:}} This is the final part of the system, where the system takes text format as an input. Then the system goes through the text to speech system and generates a speech file as an output, for this system we use eSpeakNG~\cite{duddington2012espeak}.  
  
\end{itemize}
\subsection{Result and analysis}

To get the final result which is speech based feedback, the system is mainly dependent on the input text. Therefore, image to text conversion is the main backbone of the system. If we get noisy results in the image to text then there will be no any use of going to the next step as it will always give a confusing speech based feedback to the user. To overcome this issue we compared different OCR tools to check which will provide the best result, e.g., Tesseract(which is one of the most used and supported language tools)~\cite{smith2007overview}, OCRopus~\cite{breuel2008ocropus}, OCR.space, GOCR\footnote{http://jocr.sourceforge.net/; Access Date: 30 August 2018}. Out of all these, we found Tesseract give the best result as compare to other tools. But, instead of the best result, the output of the tesseract is also very poor and noisy. It's almost impossible to proceed to the next step of the model to get the final result.  A sample of images with their corresponding result is shown in Figure~\ref{ocrRes}.

As you can see from the results, the top two images give very noisy results where the best OCR system also fails. Also, the two images placed in the bottom right corner does not provide any output which is also a failure of the system. But, the below left corner road panel an above average result which is of course not perfect but at least the system resulted some valuable information which can be an output for the second phase text to speech model.

The final speech result is always dependent upon the text input, where text input is mainly dependent on the quality of the image, focus on the text of the road panel etc. It's always difficult for a camera to capture this kind of image or detection of the best quality image during driving. As our goal is to give a speech based feedback to the driver or the system to take about the upcoming or current status of the trip, the image should detect and return the speech based feedback before the vehicle cross the point or that road panel. If the feedback delivered after the specific time or point then there will be no any use of the same. Therefore, it's necessary to keep the speed of the vehicle in a range and also with a high-quality camera that it can capture a good quality image in a specific time. Also, the current state of the art OCR techniques will be not as helpful as shown in Figure~\ref{ocrRes}.


\section{Conclusion \& Future Work}

Assisting the driver by speech is a valuable input to the intelligent transport system. In this research, we developed the same for traffic panels of four categories that contains the texts of Assamese, Hindi and English languages.   
 The designed system comprises of three parts: the first part detects text area using openCV, the second part converts the image to text and finally, the extracted text is passed through a text-to-speech generation module. Thus, the accuracy of this speech-based driving assistance system is based on accuracy of all the three modules. The results obtained from $1^{st}$ and final steps are comparable with the state-of-the-art of the existing systems. The reason behind the falling of overall accuracy was because of the state-of-the-art of the publicly available OCR tools drastically fails to recognise texts from some of the images of the developed repository. 

The accuracy of this high-end speech-based driving assistance system falls, because of the dependency in image to text conversion module. Being ongoing work, near future, we will try to enhance the accuracy of the image to text module using plug-ins or added information such as bag-of-words. 
Techniques, such as, embedding of character or word level n-grams and geographical location names may also improve the accuracy of the image-to-text module. 

\section*{Acknowledgment}

The research reported in this paper has been supported by the TEQIP III.
\bibliographystyle{IEEEtran}

\bibliography{ref}

\end{document}